\documentclass{article}

\def\mytitle#1{\setcounter{equation}{0}
\setcounter{footnote}{0}
\begin{flushleft}\Large\textbf{#1}\end{flushleft}
\vspace{0.25cm}}
\def\myname#1{\leftline{{\large #1}}\vspace{-0.13cm}}
\def\myplace#1#2{\small\begin{flushleft}\textit{#1}\\
\texttt{#2}\end{flushleft}}
\newenvironment{contribution}{\normalsize\noindent}{}
\def\myclassification#1{\small\noindent
Pacs no :
       #1\vspace{0.5cm}}
\usepackage{graphicx}
\begin{document}

\mytitle{ Emergent scenario and different anisotropic models}

\vskip0.2cm \myname{Sudeshna Mukerji\footnote{mukerjisudeshna@gmail.com}}
\vskip0.2cm
\myname{Nairwita Mazumder\footnote{nairwita15@gmail.com}}
\vskip0.2cm
\myname{Ritabrata Biswas\footnote{biswas.ritabrata@gmail.com}}
\vskip0.2cm
\myname{Subenoy Chakraborty\footnote{schakraborty@math.jdvu.ac.in}}
\vskip0.2cm

\myplace{Department of Mathematics, Jadavpur University, Kolkata-32, India.}{}

\myclassification{98.80.Cq, 98.80.-k}

\begin{abstract}
In this work, Emergent Universe scenario has been developed in general homogeneous anisotropic model and for the inhomogeneus LTB model. In the first case, it is assumed that the matter in the universe has two components - one is perfect fluid with barotropic equation of state $p=\omega\rho$ ($\omega$, a constant) and the other component is a real or phantom (or tachyonic) scalar field. In the second case, the universe is only filled with a perfect fluid and possibilities for the existence of emergent scenario has
been examined.

Keywords: Emergent Scenario, General anisotropic model, LTB model.
\end{abstract}

\begin{contribution}
\end{contribution}
\section{Introduction}

The idea of emergent universe is the result of the search for
singularity-free inflationary model in classical general
relativity.  An emergent universe model can be defined as a
singularity free universe which is ever existing with an almost
static nature in the infinite past ($t\rightarrow -\infty$) and
then evolves into an inflationary stage. In fact, an extension of
the original Lemaitre-Eddington model can be termed as emergent
universe. There are several features for the emergent universe
viz. $(i)$ the universe is almost static at the infinite past, $(ii)$
there is no timelike singularity, $(iii)$ the universe is always
large enough so that classical description of space time is
adequate $(iv)$ the universe has accelerated expansion, etc.

However, there was singularity-free solution in the literature
since 1967 - Harrison described a closed model of the universe
with radiation, which coincides with Einstein static model in
infinite past. But actual search for Emergent model of the
universe was started ( about 40 years back ) by Ellis and
collaborators (2004, 2004). They formulated a closed model of the
universe filled with two non-interacting fluids - one is a
minimally coupled scalar field having self-interaction potential
and the other is a perfect fluid with equation of state $p=\omega
\rho$. In fact,they studied only the asymptotic behaviour to
characterize the emergent scenario without finding exact analytic
solutions. Then Mukherjee et. al. (2005) solved semi-classical
equations in the Starobinsky model for flat FRW space-time and
examined the features of emergent scenario. Subsequently,
Mukherjee and others (2006) were able to obtain nonsingular (i.e.
geodesically complete) inflationary solution where a part of the
matter is in exotic form. Mulryne et. al. (2005)  have discussed the
existence and stability of emergent models in the context of Loop Quantum Cosmology.  Debnath (2008) has formulated
an emergent model of the universe for exotic matter in the form
of phantom or tachyonic field. Banerjee et. al. (2007, 2008) have shown a
model of emergent universe in brane scenario while Campo et.al.
(2007)have studied a model of emergent universe for self-
interacting Brans-Dicke theory. Recently, Mukerji et.al. (2010, 2010) have formulated an
emergent universe model in the context of Einstein-Gauss-Bonnet
theory and in Horava Gravity.

 In section 2, we consider a homogeneous and anisotropic space time models
described by the line element
\begin{equation}\label{1}
ds^{2} = -dt^{2}+ a^{2}dx^{2}+ b^{2}d\Omega_{k}^{2}
\end{equation}
where the scale factors $a$ and $b$ are functions of time $t$
alone. We note that
$$d\Omega_{k}^{2}= dy^{2}+ dz^{2},~~~  when~~ k~ =~ 0 ~~ (Bianchi~
I~
model)$$
 $$~~~~~~~~~~~~~~~~~~~~~~~~~~~~ = d\theta^{2} + \sin^{2}\theta d\phi^{2},~~when~~ k ~=
 ~+1~~~
(Kantowaski-Sachs ~model)$$\
 $$~~~~~~~~~~~~~~~~~~~~ = d\theta^{2} + \sinh^{2}\theta d\phi^{2},~~when ~~k~ =~ -1~~~(Bianchi
 ~III~
model)$$
Here $k$ is the scalar curvature and the above three types are
characterized by Thorne (1967)  as flat, closed and open
respectively.

Now, the expression for the Hubble parameter $H$, the deceleration
parameter $q$ and the anisotropy scalar $\sigma$ in terms of the scale factors are :
$$~~H~ = ~\frac{1}{3}\left[\frac{\dot{a}}{a}+ ~2~
\frac{\dot{b}}{b}\right]$$
$$ ~q~=~ -1~ -~\frac{\dot{H}}{H^{2}}$$and $$\sigma^{2}=\frac{1}{3}\left(\frac{\dot{a}}{a}-\frac{\dot{b}}{b}\right)^{2}$$\\

In section 3, we consider our universe as inhomogeneous
$Lema\hat{i}tre-Tolman-Bondi [LTB] Model$. The LTB model is one of the most wellknown sphercally symmetric models in general relativity. It was described by Lemaitre (1933), Tolman (1934) and Bondi (1947) during the period of time from 1933 to 1947. The exact solution have been obtained by Bonnor (1972, 1974). This simple
inhomogeneous cosmological model agrees with current supernova and
some other data (Moffat 2005; Moffat 2006; Alnes 2006; Alnes 2007; Romano 2010, 2010). Also very recently  Clarkson and Marteens (2010)
give a justification for inhomogeneous model from the point of view
of perturbation analysis. 

The inhomogeneous spherically symmetric LTB space-time model is described by the metric ansatz in a co-moving frame as
\begin{equation}\label{2}
ds^{2}=-dt^{2}+\frac{R'^{2}}{1+f(r)}dr^{2}+R^{2}\left(d\theta^{2}+\sin^{2}\theta d\phi^{2}\right)
\end{equation}
where $R=R(r,t)$ is the area radius of the spherical surfaces and $f(r)(>-1)$ is the curvature scalar that classifies the space-time as

(i) bounded, if, $-1<f(r)<0$,

(ii)marginally bounded, if, $f(r)=0$,

(iii)unbounded if $f(r)>0$.

The energy-momentum tensor for perfect fluid is given by
\begin{equation}\label{3}
T_{\mu\nu}=\left(\rho+p\right)u_{\mu}u_{\nu}+pg_{\mu\nu}
\end{equation}
where the fluid 4-velocity $u^{\mu}$ has normalization $u^{\mu}u_{\mu}=-1$ and $\rho$, $p$ are respectively the usual matter density and pressure of the fluid. Finally, the paper ends with a brief discussion in section 4.
\section{General anisotropic model : Basic Equations for Emergent Scenario
:}

The Einstein field equations for homogeneous and anisotropic model of
the universe  with non-interacting two fluid system can be written as :
\begin{equation}\label{4}
\frac{\ddot{a}}{a} + 2 \frac{\ddot{b}}{b} =
\frac{-1}{2}[ \rho_{m} + \rho_{\phi} + 3 p_{m} + 3p_{\phi}]
\end{equation}
and
\begin{equation}\label{5}
\frac{\dot{b}^{2}}{b^{2}} + 2 \frac{\dot{a}}{a} \frac{\dot{b}}{b} +\frac{k}{b^{2}} =[ \rho_{m} +\rho_{\phi}]
\end{equation}
where $\rho_{m}$ and $p_{m}$ are the energy density and pressure
of a perfect fluid having equation of state $p_{m}=\omega\rho_{m}$,
($\omega$,a constant), $\rho_{\phi}$ and $p_{\phi}$ are the
energy-density and pressure of a scalar field $\phi$ having
expressions

\begin{equation}\label{6}
\rho_{\phi} = \frac{\varepsilon}{2} \dot{\phi}^{2} + V(\phi)
\end{equation}

\begin{equation}\label{7}
p_{\phi} = \frac{\varepsilon}{2} \dot{\phi}^{2} - V(\phi)
\end{equation}

Here $V(\phi)$ is the potential for the scalar field $\phi$ and
$\varepsilon =\pm1$ corresponds to normal or phantom scalar field.

As the two components of the matter field are
non-interacting, so the energy conservation equations are
given by
\begin{equation}\label{8}
\dot{\rho_{m}}+ 3H (\rho_{m} + p_{m})=0
\end{equation}
and
\begin{equation}\label{9}
\dot{\rho}_{\phi}+3H(\rho_{\phi}+p_{\phi})=0
\end{equation}

For emergent scenario the appropriate form of the scale factors
will be according to Debnath (2008)
\begin{equation}\label{10}
a(t)=a_{0}[ \beta + e^{\alpha t}]^{n}
\end{equation}
and
\begin{equation}\label{11}
b(t)=b_{0}[ \delta + e^{\gamma t}]^{m}
\end{equation}
where $a_{0}$ , $ \beta $ ,$ \alpha $ , $ n $ and $b_{0}$ ,
$ \delta $ ,$ \gamma $ , $ m $  are positive
constants. The justification for such  choices are the following :

1. $a_{0}>0$ and $b_{0}>0$ for scale factor to be positive
definite.

2. $\beta>0$ and $\delta>0$, otherwise there will be big rip
singularity.

3. $\alpha>0$, $n>0$, $\gamma>0$ and $m>0$  for expanding model of
the universe.

4.$\alpha<0$, $n<0$ or $\gamma<0$, $m<0$ implies that there was a
big bang singularity at infinite past.

For the above choice of the scale factors, the universe started
with a finite volume at $t=-\infty$, grows gradually without
encountering any singularity for any $t$ and finally, the
universe will be of infinite volume at future infinity.\\

\begin{figure}
\includegraphics[height=4in, width=4in]{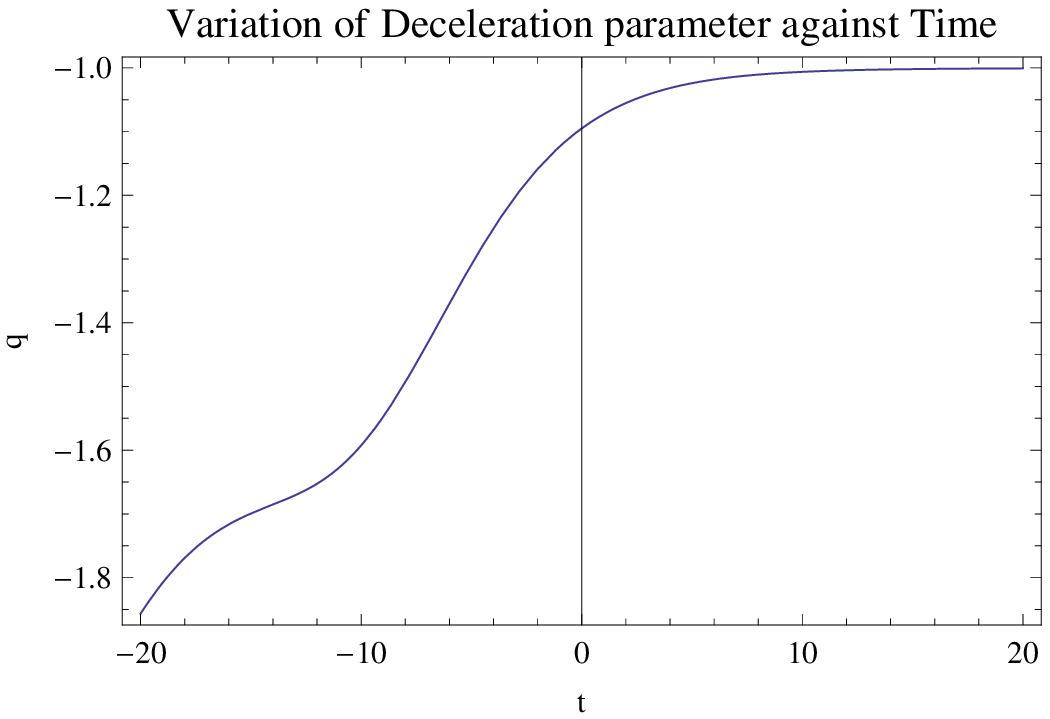}\\
\vspace{1mm} ~~~~~~~~~~~~~~~~~~~~~~~~~~~~~~~~~Fig.1~~~~~~~~~~~~~~~~~~~~~~~~~~~~~\\

\vspace{6mm} Fig. 1 represents the graph of the deceleration parameter, $q$ against the cosmic time,$t$, in anisotropic model of universe for
the constants $\alpha=0.1,~\beta=0.2,~\gamma=0.3,
~\delta=0.4,~~m=4,~n=6$.

\vspace{6mm}

\end{figure}

For this choice of the scale factors $'a'$ and $'b'$, the Hubble
parameter, it's derivatives and the deceleration parameter $'q'$
have the following expressions :

$$H = \frac{1}{3}\left[\frac{[n \alpha e^{\alpha  t}]}
      {[\beta + e^{\alpha  t}]} + \frac{[2m \gamma e^{\gamma  t}]}
      {[\delta + e^{\gamma  t}]}\right]~,$$

$$\dot{H}= \frac{1}{3}\left[\frac{[n\beta (\alpha^{2})e^{\alpha  t}]}
         {[\beta+ e^{\alpha  t}]^{2}} + \frac{[2m \delta (\gamma^{2})e^{\gamma  t}]}
         {[\delta+ e^{\gamma  t}]^{2}}\right]~,$$

$$\ddot{H}= \frac{1}{3}\left[\frac{[n \beta \alpha^{3}e^{\alpha  t}
           (\beta - e^{\alpha  t})]}
           {[\beta + e^{\alpha  t}]^{3}} + \frac{[2m \delta \gamma^{3}e^{\gamma  t}
           (\delta - e^{\gamma  t})]}
           {[\delta + e^{\gamma  t}]^{3}}\right]~,$$

\begin{equation}\label{12}
q = -1 - \frac{3 n \beta \alpha^{2}e^{\alpha t} (\delta +
e^{\gamma t})^{2} +6 m \delta \gamma^{2}e^{\gamma t}(\beta
+e^{\alpha t})^{2} }{ [n  \alpha e^{\alpha t} (\delta + e^{\gamma
t}) +2 m \gamma e^{\gamma t}(\beta +e^{\alpha t})]^{2}}
\end{equation}

We note that $H$ and $\dot{H}$ are positive definite while $q$ is
negative definite throughout the evolution but $\ddot{H}= 0$ when
$t$ satisfies
$$\frac{(\delta ~+e ^{\gamma t})^{3} ~(\beta - e^{\alpha t})~ e^{\gamma t}}{(\beta ~+e ^{\alpha t})^{3} ~(e^{\gamma t}- \delta)~ e^{\gamma t}}
 = \frac{2 m \delta\gamma^{3}}{n\beta\alpha^{3}}.$$
Asymptotically, as
$t\rightarrow -\infty$, $H$, $\dot{H}$, and $\ddot{H}$ all tend
to zero but $q\rightarrow -\infty$ while the model becomes a de
Sitter universe as $t\rightarrow \infty$

In figure 1 we have plotted the deceleration parameter, $q$, against time ,$t$. As $\dot{H}$ is positive throughout the evolution so $q$ is always less than $-1$ while asymptotically as $t \rightarrow +\infty$, $\dot{H}\rightarrow 0$ and hence from (\ref{12}), $q\rightarrow -1$ as shown in figure 1.

Now using the equation of state for the perfect fluid the energy
conservation relation (\ref{8}) can be integrated to obtain
\begin{equation}\label{13}
\rho_{m} = \rho_{0}(ab^{2})^{-(1+ \omega )},
\end{equation}
where $\rho_{0}$ is an integration constant.

Also, writing the explicit form of $\rho_{\phi}$ and $p_{\phi}$ ,
from the field equations (\ref{4}) and (\ref{5})

\begin{equation}\label{14}
\rho_{\phi}= \frac{\dot{b}^{2}}{b^{2}} + 2 \frac{\dot{a}}{a}
\frac{\dot{b}}{b} +\frac{k}{b^{2}} -\rho_{m}
\end{equation}

and

\begin{equation}\label{15}
p_{\phi}=-\frac{1}{3}\left[\frac{2\ddot{a}}{a}+\frac{4\ddot{b}}{b}
+\frac{\dot{b}^{2}}{b^{2}}+
\frac{2\dot{a}\dot{b}}{ab}+\frac{k}{b^{2}}\right]-\omega \rho_{m}
\end{equation}

i.e.,  we have
\begin{equation}\label{16}
 \dot{\phi}^{2}= \frac{1}{\varepsilon}\left[\frac{2}{3}\left[\frac{k}{b^2}-3\dot{H}-3{\sigma}^2\right]
  -(1 + \omega) \rho_{0}(ab^{2})^{-(1 + \omega)}\right]
\end{equation}

and

\begin{equation}\label{17}
  V(\phi) = \left[\dot{H} + 3H^2+\frac{2k}{3b^2}\right]- \frac{1}{2}(1-\omega ) \rho_{0}(ab^{2})^{-(1 + \omega)}
\end{equation}

We note that $V(\phi)$ is independent of $\varepsilon$ i.e. it has
the same value for real or phantom scalar field.

\subsection{Possibility of Emergent Scenario: The
restrictions.}

In this section, we discuss the possibility of emergent universe
and present the restrictions for  its validity. In other words,
we examine whether the choices of $a(t)$ and $b(t)$  given by
equations (\ref{10})and (\ref{11}) are possible solution for the field
equations for (i)the real scalar field and (ii) the phantom scalar field.\\

\subsubsection{Real Scalar field ($\varepsilon=+1$) :}

      In this case, the expression for $ \dot{\phi}^{2}$ is

$$\dot{\phi}^{2} =\frac{2}{3}\left[\frac{k}{b^2}-3\dot{H}-3 {\sigma}^2\right]
  -(1 + \omega) \rho_{0}(ab^{2})^{-(1 + \omega)}$$

  When the perfect fluid matter component is not of exotic
  nature (i.e. does not violate the weak energy condition) then
  for real scalar field, emergent scenario is not possible for
  $k=0,-1$(i.e. for flat and open model) while for Kantawski-Sachs
  model (with $k=+1$),emergent scenaeio is possible provided

\begin{equation}\label{17a}
\frac{1}{b^2}>3\dot{H}-3 {\sigma}^2+\frac{3}2 (1+\omega)
{\rho}_0 {(ab^2)}^{-(1+\omega)}
\end{equation}

On the other hand, if the perfect fluid matter component is of
phantom nature, then emergent scenario is possible for all three
models provided

\begin{equation}\label{17b}
|1+\omega|{\rho}_0{(ab^2)}^{-(1+\omega)}>3\dot{H}+3
{\sigma}^2-\frac{k}{b^2}
\end{equation}

 and $\phi$ can be obtained as
 
 $$\phi = \int\sqrt{ \frac{2}{3}\left[\frac{\dot{b}^{2}}{b^{2}}
  + \frac{2 \dot{a} \dot{b}}{a b}- \frac{\ddot{a}}{a} - \frac{2 \ddot{b}}{b}+
  \frac{k}{b^{2}}\right]
  -(1 + \omega) \rho_{0}(ab^{2})^{-(1 + \omega)}}dt$$


      As $a$ and $b$ are given functions of $t$, (equation(\ref{10}) and (\ref{11})), so in principle
 the above integral can be evaluated to obtain $\phi$ as a function of $t$.

\begin{figure}
\includegraphics[height=1in, width=1in]{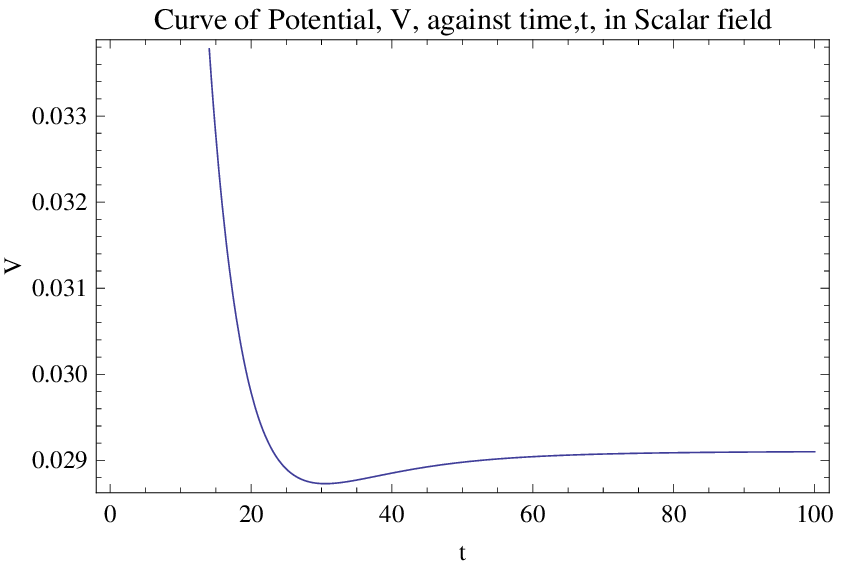}~~~~
\includegraphics[height=1in, width=1in]{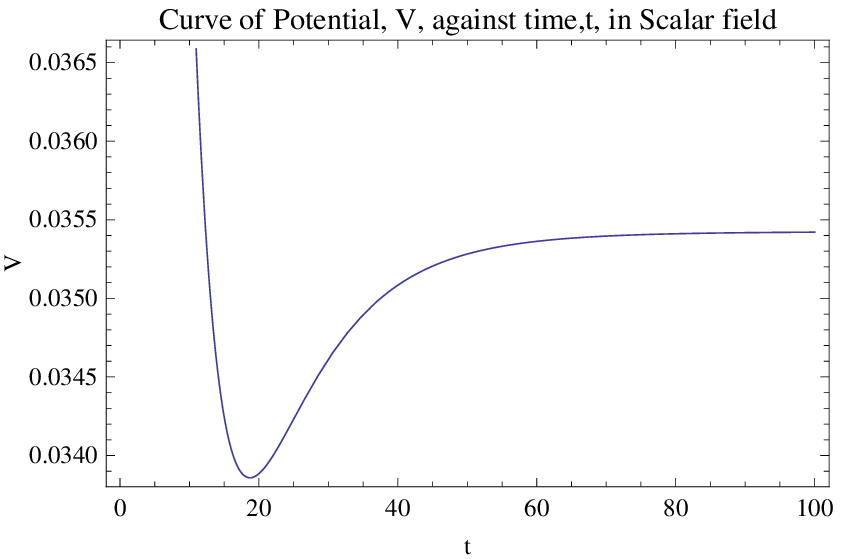}~~~~
\includegraphics[height=1in, width=1in]{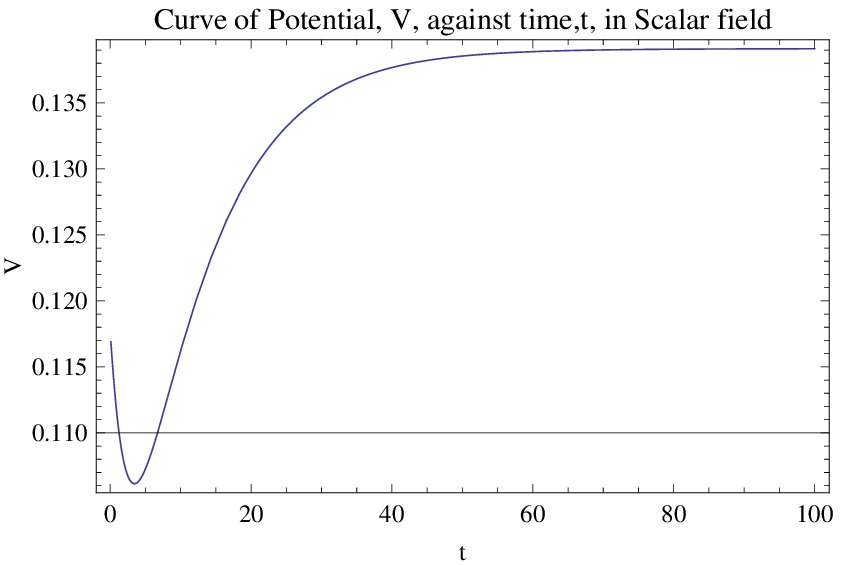}~~~~
\includegraphics[height=1in, width=1in]{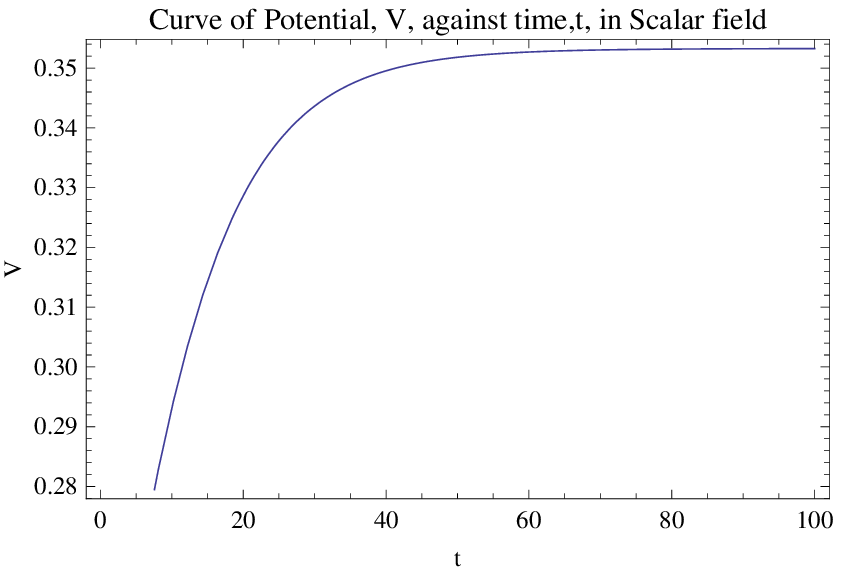}\\
\vspace{1mm} ~~~~~~~~~~~~Fig.2a~~~~~~~~~~~~~~~~Fig2b~~~~~~~~~~~~~~~~~~~~~Fig.2c~~~~~~~~~~~~~Fig.2d~~~~~~~~\\
\includegraphics[height=1in, width=1in]{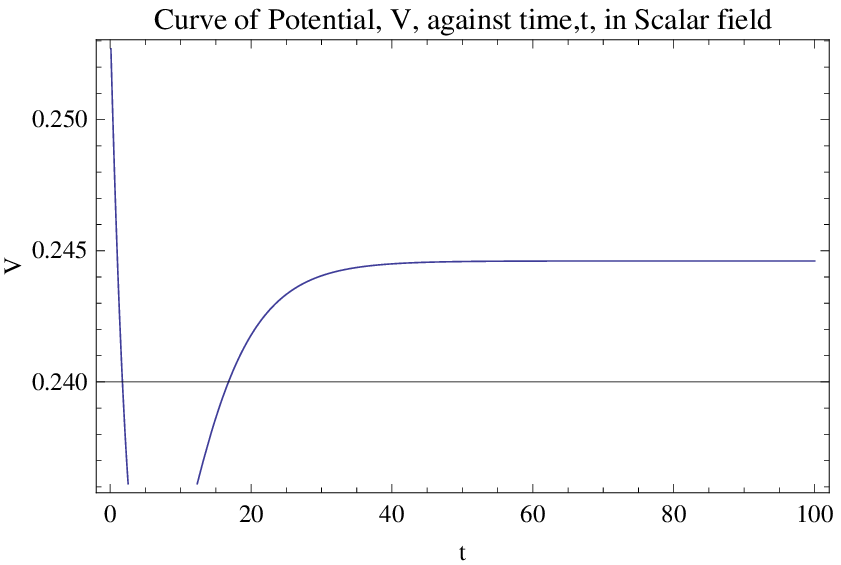}~~~~
\includegraphics[height=1in, width=1in]{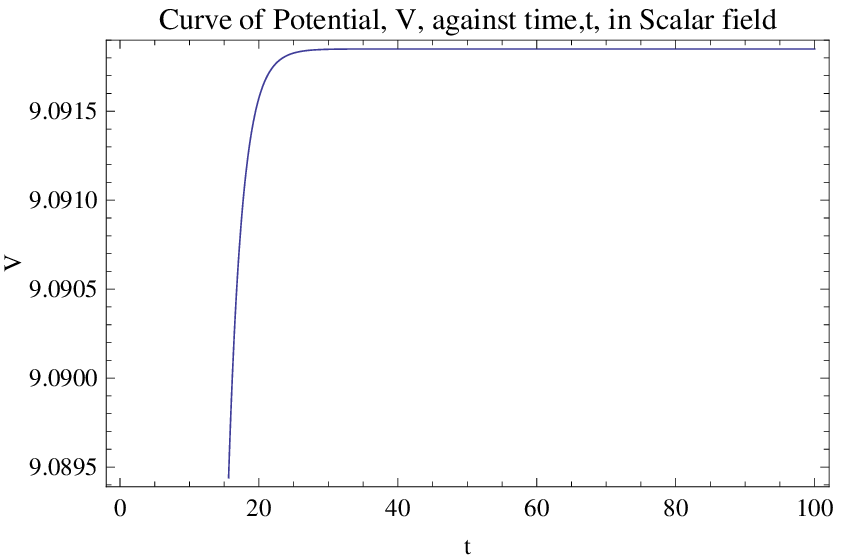}~~~~
\includegraphics[height=1in, width=1in]{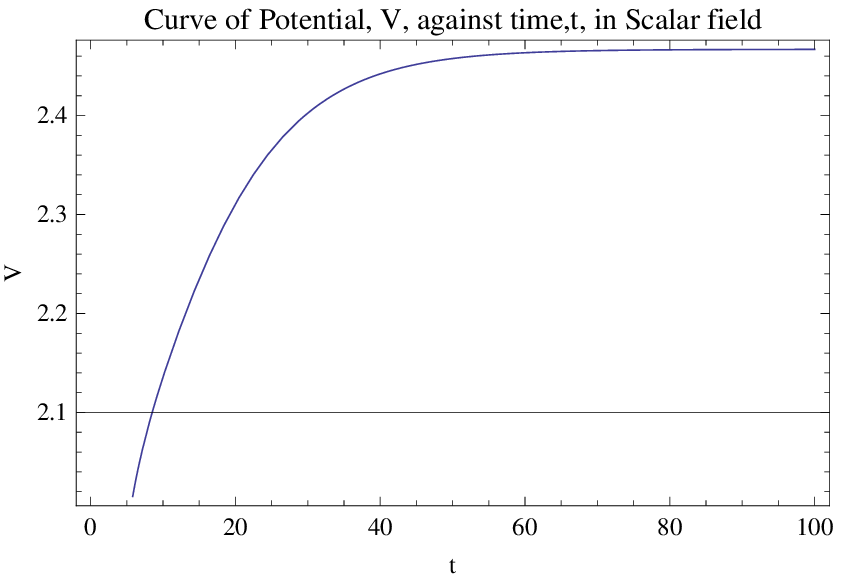}~~~~
\includegraphics[height=1in, width=1in]{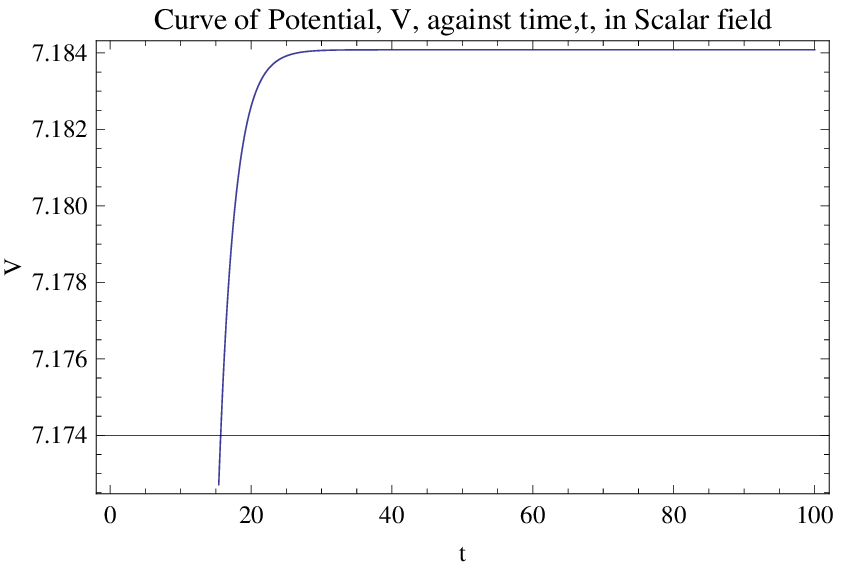}\\
\vspace{1mm} ~~~~~~~~~~~~Fig.2e~~~~~~~~~~~~~~~~Fig2f~~~~~~~~~~~~~~~~~~~~~Fig2g~~~~~~~~~~~~~Fig.2h~~~~~~~~\\
\vspace{1mm} Fig. 2a represents the graph of the scalar field potential $V$ against cosmic time,$t$,for the variables $\alpha=0.05,~\beta=0.05,~\gamma=0.1,~\delta=0.3,~~m=1.05,~n=1.71,~a_{0}=1,~b_{0}=2,~\rho_{0}=1,~\omega=1/3$ at k=1,i.e., for the closed universe.Fig. 2b represents the graph  for the same value of the variables except for$ ~~m=1.38,~n=1.$ Fig. 2c represents the graph  for the same value of the variables except for $~~m=2.98.$ Fig. 2d represents the graph  for the same value of the variables except for $~m=4.72,~n=1.71$
Fig. 2e represents the graph  for the same value of the variables except for $~m=1,~n=2.48,\alpha=0.218,~\beta=0.228,~\gamma=0.158,~\delta=0.3419,$and a very low value of $~\rho_{0}=0.001$
 Fig. 2f represents the graph of the scalar field potential $V$ against cosmic time,$t$,for the variabless $\alpha=0.812,~\beta=0.456,~\gamma=0.496,~\delta=0.596,~~m=2.85,~n=2.95,$ at k=0,i.e., for the flat universe.Fig. 2g represents the graph of the scalar field potential $V$ against cosmic time,$t$,for the variables $\alpha=0.608,~\beta=0.336,~\gamma=0.1,~\delta=1.39,~~m=2.81,~n=3.55,$ at k=-1,i.e., for the open universe..Fig. 2h represents the graph of the scalar field potential $V$ against cosmic time,$t$,for the same value of the variables  except $\gamma=0.442$ at k=-1,i.e., for the open universe.
\vspace{1mm}
\vspace{6mm}
\end{figure}

The graphical representation of $V$ for various choices of the parameters and the curvature scalar $k$ are shown in figure fig 2(a)-2(h). The first five figures are for $k=+1$ then next two are for $k=-1$ and the last one is for $k=0$. The curves for the potential show some distinct features for different choices of the parameters.



\subsubsection{Phantom Scalar field $(\varepsilon=-1)$ :}
In this case, $\dot{\phi}^{2}$ can be written as

  $$\dot{\phi}^{2} =\frac{2}{3}\left[3\dot{H}+3 {\sigma}^2 -\frac{k}{b^2}\right]
  +(1 + \omega) \rho_{0}(ab^{2})^{-(1 + \omega)} $$
when the perfect fluid matter component is not of exotic nature,
the emergent scenario is possible for phantom scalar field when
$k=0,-1$ (i.e. Bianchi I and Bianchi III model) while for
Kantowski Sachs model (k=1) emergent scenario is possible provided

\begin{equation}\label{17c}
\frac{1}{b^2}<3\dot{H}+3 {\sigma}^2+\frac{3}2 (1+\omega)
{\rho}_0 {(ab^2)}^{-(1+\omega)}
\end{equation}

But when the perfect fluid matter component is of phantom nature,
then emergent scenario is always possible for all three models provided

\begin{equation}\label{17d}
|1+\omega|{\rho}_0{(ab^2)}^{-(1+\omega)}<3\dot{H}+3
{\sigma}^2-\frac{k}{b^2}
\end{equation}

\subsection{Emergent Scenario for Tachyonic Scalar field : }

For a tachyonic scalar field $\psi$, with potential $B(\psi)$,
the energy density $\rho_{\psi}$ and pressure $p_{\psi}$ are
given by

\begin{equation}\label{20}
\rho_{\psi} = \frac{B(\psi)}{\sqrt{1 - \varepsilon \dot{\psi}^{2}}}
\end{equation}
and
\begin{equation}\label{21}
p_{\psi} = -B(\psi) \sqrt{1 - \varepsilon \dot{\psi}^{2}}
\end{equation}
    where as before $\varepsilon = \pm1$ corresponds to normal and
phantom tachyonic field. So, $\dot{\psi}^{2}$ and  $B(\psi)$ can
be written as

\begin{equation}\label{22}
  \dot{\psi}^{2} = \frac{\rho_{\psi}+ p_{\psi}}{ \varepsilon (\rho_{\psi})}
\end{equation}
and
\begin{equation}\label{23}
  B(\psi) = \sqrt{- \rho_{\psi} p_{\psi}}
\end{equation}

Hence from the filed equations $\dot{\psi}^{2}$ and $B(\psi)$ have the expressions
\begin{equation}\label{24}
  \dot{\psi}^{2} = \left(\frac{1}{\varepsilon}\right)\frac{\frac{2}{3}\left[\frac{k}{b^2}-3\dot{H}-3 {\sigma}^2\right]
  -(1 + \omega) \rho_{0}(ab^{2})^{-(1 + \omega)} }
                  {\frac{\dot{b}^{2}}{b^{2}} + \frac{2 \dot{a} \dot{b}}{a b} + \frac{k}{b^{2}}
                  - \rho_{0} (a b^{2})^{-(1 + \omega)}}
\end{equation}

and

\begin{equation}\label{25}
   B(\psi) = \sqrt{\frac{2}{3}\left[\frac{\dot{a}^{2}}{a^{2}}
  + \frac{ 5{\dot{b}}^2 }{ 2b^2}+ \frac{\dot{a}\dot{b}}{ab} +
  3\dot{H}+ \frac{k}{b^{2}}\right]
  + \omega \rho_{0}(ab^{2})^{-(1 + \omega)}} \times \sqrt{\frac{\dot{b}^{2}}{b^{2}} + \frac{2 \dot{a} \dot{b}}{a b} + \frac{k}{b^{2}}
                  - \rho_{0} (a b^{2})^{-(1 + \omega)}}
\end{equation}
We note that as before $B(\psi)$ is independent of $\varepsilon$ i.e. it has
the same value for normal or phantom tachyonic field.

\begin{figure}
\includegraphics[height=1in, width=1in]{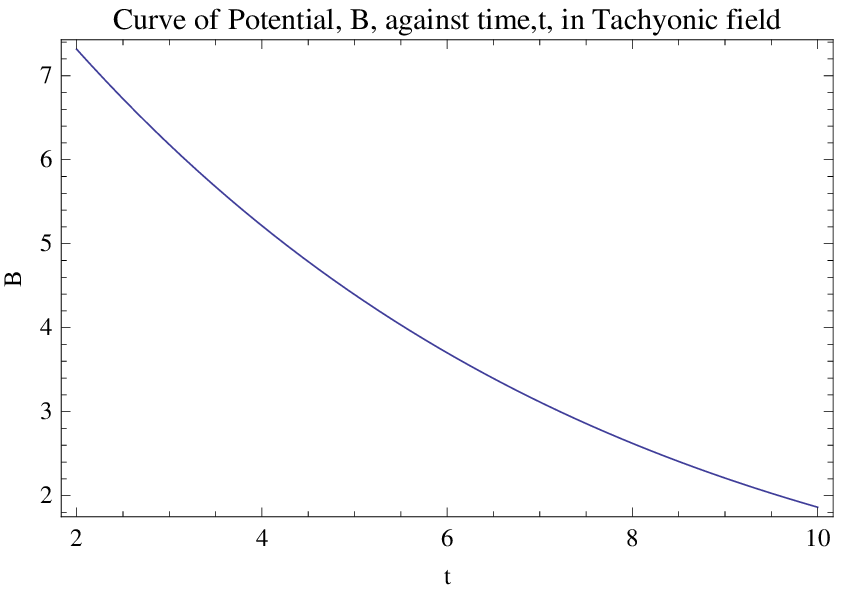}~~~~
\includegraphics[height=1in, width=1in]{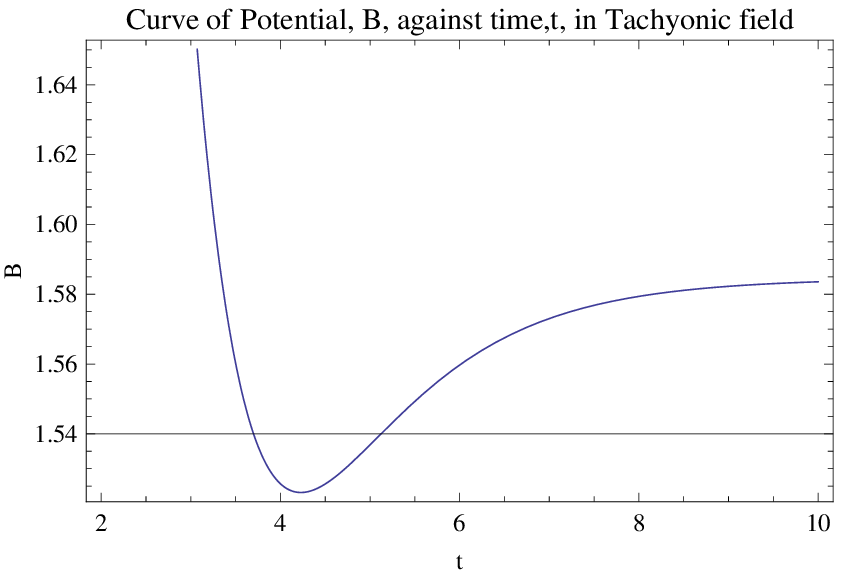}~~~~
\includegraphics[height=1in, width=1in]{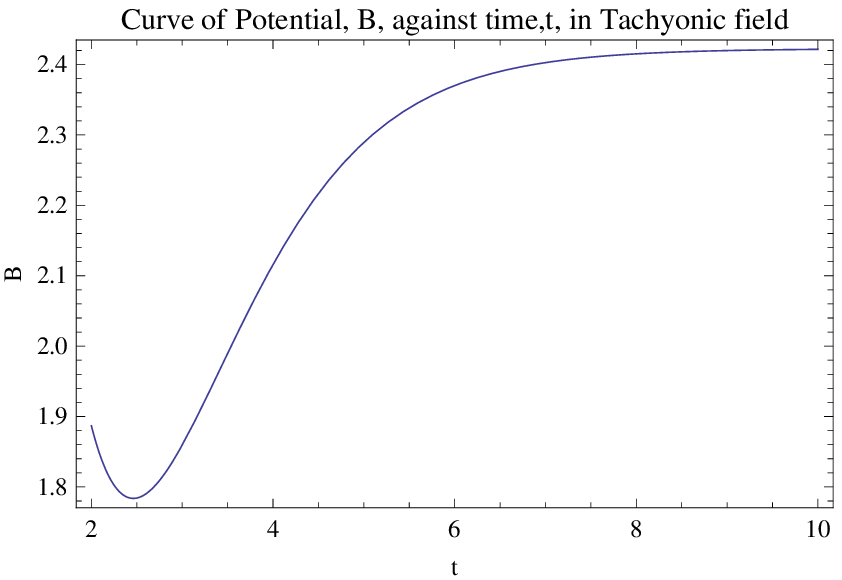}~~~~
\includegraphics[height=1in, width=1in]{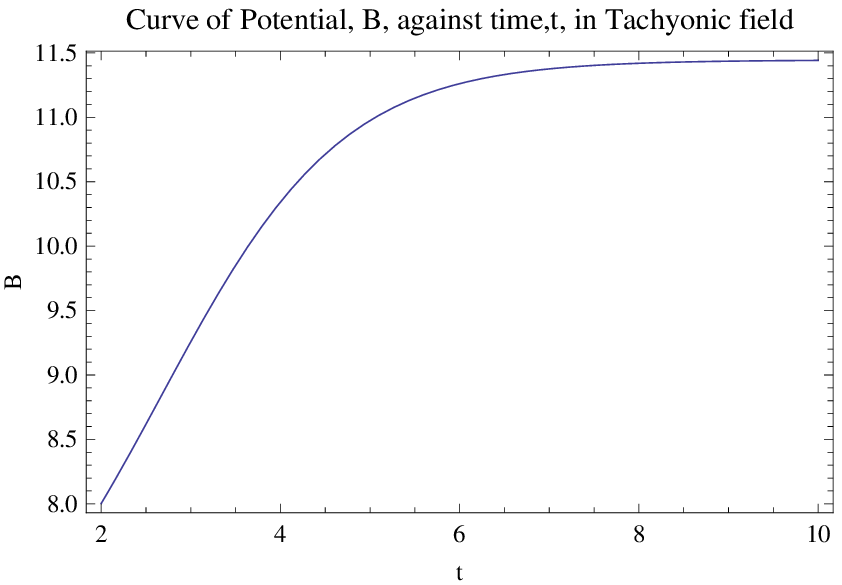}\\
\vspace{1mm} ~~~~~~~~~~~~Fig.3a~~~~~~~~~~~~~~~~Fig3b~~~~~~~~~~~~~~~~~~~~~Fig3c~~~~~~~~~~~~~Fig.3d~~~~~~~~\\
\includegraphics[height=1in, width=1in]{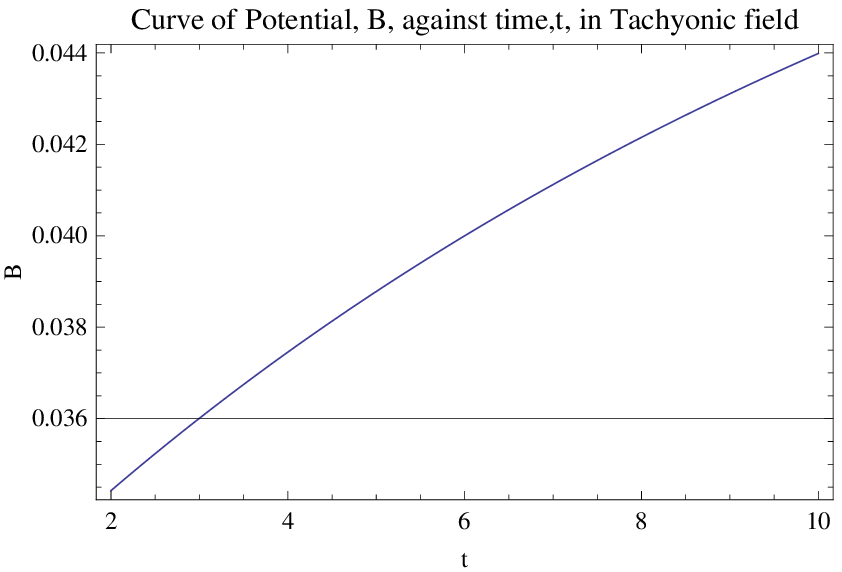}~~~~
\includegraphics[height=1in, width=1in]{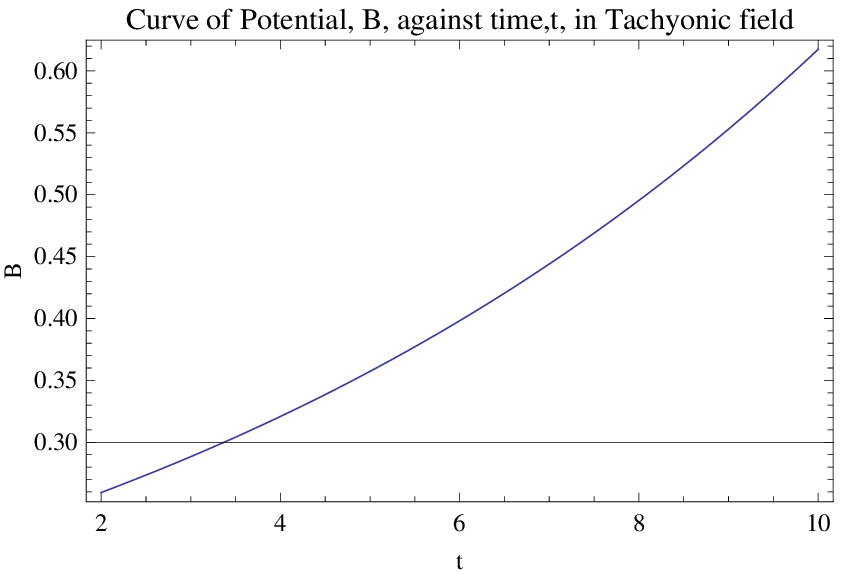}~~~~
\includegraphics[height=1in, width=1in]{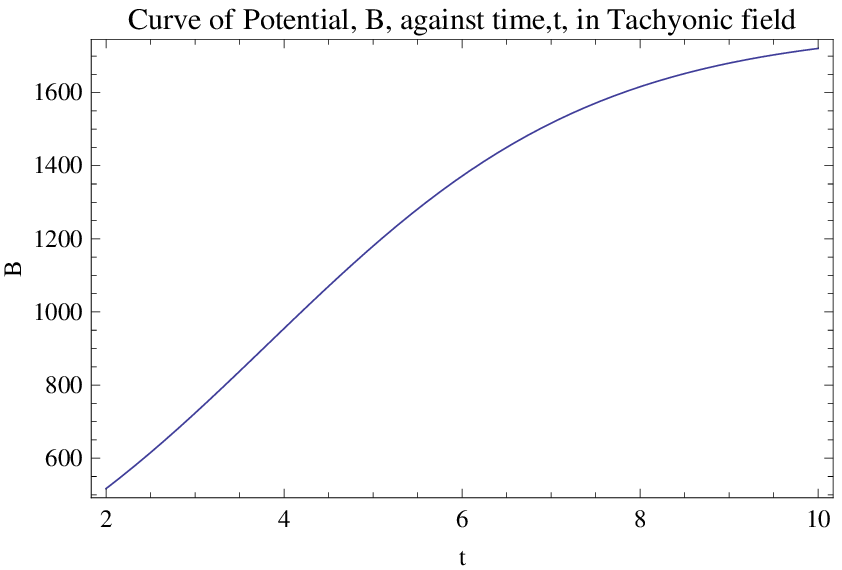}~~~~
\includegraphics[height=1in, width=1in]{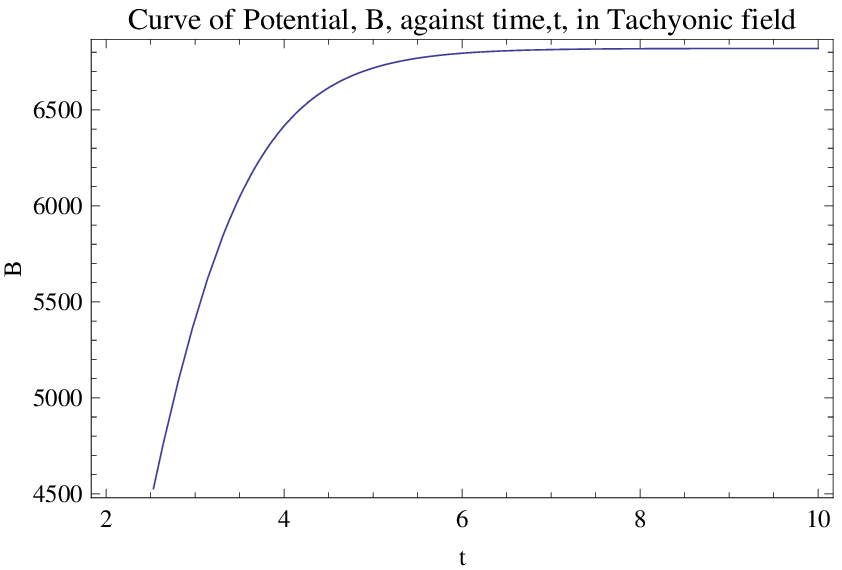}\\
\vspace{1mm} ~~~~~~~~~~~~Fig.3e~~~~~~~~~~~~~~~~Fig3f~~~~~~~~~~~~~~~~~~~~~Fig3g~~~~~~~~~~~~~Fig.3h~~~~~~~~\\
\vspace{1mm} Fig. 3a represents the graph of the tachyonic field potential $B$ against cosmic time,$t$,for the constants $\alpha=0.1,~\beta=0.2,~\gamma=0.1,~\delta=0.2,~~m=1,~n=1,~a_{0}=0.2,~b_{0}=0.2,~\rho_{0}=0.01,~\omega=-1$ at k=1,i.e., for the closed universe.Fig. 3b represents the graph of the tachyonic field potential $B$ against cosmic time,$t$,for the constants $\alpha=1,~\beta=16.2,~\gamma=0.59~\delta=0.2,~~m=1,~n=1,~a_{0}=0.2,~b_{0}=0.2,~\rho_{0}=0.01,~\omega=-1$ at k=1,i.e., for the closed universe. In 3c and 3d all the data are same with 3b except for $\gamma$ which has values respectively 0.92 and 3.51 in these two cases.
3e and 3f represents the graph of the tachyonic field potential $B$ against cosmic time $t$ at k=-1,i.e. open universe and 3g-3h represents the tachyonic field potential $B$ at k=0, i.e. flat universe.

\vspace{1mm}
\vspace{6mm}
\end{figure}

The variation of $B$ with cosmic time $t$ for different choices of the parameters and the curvature scalar $k$ are shown in fig. 3(a)-(h). The first four figures are for $k=1$,then two for $k=-1$ and the last two for $k=0$. In this case also the curves for the potential show some special feature for different choices of the parameters.

\subsubsection{Normal Tachyonic Scalar field
$(\varepsilon=+1)$ :}

  Here for real $\psi$ both numerator and denominator in
equation (\ref{24}) must have the same sign. The denominator cannot be
negative because then the expression within the second square root
in equation (\ref{25}) becomes negative and hence $B(\psi)$ becomes
imaginary. So,the numerator has to be positive. Hence, when the perfect fluid matter component is not of exotic nature (i.e., does not violate the weak energy condition) then emergent scenario is not possible in this case for flat Bianchi $I$ ($k=0$)model and open Bianchi $III$ ($k=-1$) model whereas for closed Kantowaski Sachs model($k=+1$) emergent scenario is possible provided inequality (\ref{17a}) holds. When the perfect fluid matter component is of phantom nature then emergent scenario is possible for all three models provided restriction (\ref{17b}) holds.

Also $\psi$ can be obtained as
$$ \psi = \int\left[\frac{\frac{2}{3}\left\{\frac{k}{b^2}-3\dot{H}-3 {\sigma}^2\right\}
  -(1 + \omega) \rho_{0}(ab^{2})^{-(1 + \omega)} }
                  {\frac{\dot{b}^{2}}{b^{2}} + \frac{2 \dot{a} \dot{b}}{a b} + \frac{k}{b^{2}}
                  - \rho_{0} (a b^{2})^{-(1 + \omega)}}\right]^{\frac{1}{2}}dt$$

\subsubsection{Phantom Tachyonic Scalar Field $(\varepsilon=-1)$ :
}

In this case, for real $\psi$ the numerator and the denominator of equation (\ref{24}) should have opposite signs. But the denominator can not be negative as stated earlier. So, the only possibility is that the numerator must be negative definite. In this case, $\dot{\psi}^{2}$ can be written as
$$ \dot{\psi}^{2} = \frac{\frac{2}{3}\left\{3\dot{H}+3 {\sigma}^2-\frac{k}{b^2}\right\}
  +(1 + \omega) \rho_{0}(ab^{2})^{-(1 + \omega)} }
                  {\frac{\dot{b}^{2}}{b^{2}} + \frac{2 \dot{a} \dot{b}}{a b} + \frac{k}{b^{2}}
                  - \rho_{0} (a b^{2})^{-(1 + \omega)}}$$
                  So, when the perfect fluid matter component is not of exotic nature then emergent scenario is always possible when $k=0,~-1$. While for $k=+1$, emergent scenario is possible when restriction (\ref{17c}) holds. When the perfect fluid matter component is of phantom nature, then emergent scenario is possible for flat, closed and open models provided restriction (\ref{17d}) holds.

\section{Lemaitre-Tolman-Bondi model}

By introducing
the mass function $F(r,t)$ (Joshi 1979; Banerjee 2003; Debnath 2003,2004) (related to the mass contained
within the co-moving radius $r$)as
\begin{equation}\label{26}
F(r,t)=R({\dot{R}}^2-f(r))
\end{equation}
the Einstein's equations in LTB model are
\begin{equation}\label{27}
8\pi G \rho =\frac{F'(r, t)}{R^{2}R'}
\end{equation}
and
\begin{equation}\label{28}
8\pi G p =-\frac{\dot{F}(r, t)}{R^{2}\dot{R}}
\end{equation}
and the evolution equation for $R$ is
\begin{equation}\label{29}
2R\ddot{R}+\dot{R}^{2}+8 \pi G p R^{2}=f(r)
\end{equation}
The conservation equation is
\begin{equation}\label{30}
\dot{\rho}+3H\left(\rho+p\right)=0
\end{equation}
where
\begin{equation}\label{33}
H=\frac{1}{3}\left(\frac{\dot{R}'}{R'}+\frac{2\dot{R}}{R}\right)
\end{equation}
is the Hubble parameter.
The equation of state is given by
\begin{equation}\label{31}
p=\epsilon \rho ,~~ where~~\epsilon ~~is~~a~~constant.
\end{equation}
Using equations (\ref{33}) and (\ref{31}), equation (\ref{30}) can be integrated to give
\begin{equation}\label{34}
\rho=\rho_{0}\left(R^{2}R'\right)^{-\left(1+\epsilon\right)}
\end{equation}
with $\rho_{0}$, a constant of integration.

Using (\ref{31}) and (\ref{34}) in equation (\ref{29}), the differential equation for $R$ is obtained as
\begin{equation}\label{35}
2R\ddot{R}+\dot{R}^{2}+\frac{8\pi G \epsilon \rho_{0}}{\left(R^{2}R'\right)^{\left(1+\epsilon\right)}}=f(r)
\end{equation}
Now, taking $8\pi G =1$ and writing
\begin{equation}\label{36}
R=h(r)g(t)
\end{equation}
we can write (\ref{35}) as
\begin{equation}\label{37}
2g(t)\left(g(t)\right)^{3\left(1+\epsilon\right)}\frac{d^{2}}{dt^{2}}{g(t)}+\left(\frac{d}{dt}g(t)\right)^{2}\left(g(t)\right)^{3\left(1+\epsilon\right)}
+\frac{\epsilon\rho_{0}}{\left(h(r)\right)^{2}\left(\left(h(r)\right)^{2}\frac{d}{dr}h(r)\right)^{\left(1+\epsilon\right)}}=\frac{f(r)}{h^{2}}\left(g(t)\right)^{3\left(1+\epsilon\right)}
\end{equation}
Choosing $\epsilon=-1$ (i.e., matter as a cosmological constant)
the above differential equation can be written in separable
form as
$$2g(t)\frac{d^{2}}{dt^{2}}g(t)+\left(\frac{d}{dt}g(t)\right)^{2}=\frac{1}{\left(h(r)\right)^{2}}\left\{\rho_{0}+f(r)\right\}=\lambda~~~~(say)$$
i.e.,
\begin{equation}\label{38}
2g(t)\frac{d^{2}}{dt^{2}}g(t)+\left(\frac{d}{dt}g(t)\right)^{2}=\lambda
\end{equation}
and
\begin{equation}\label{39}
\frac{1}{\left(h(r)\right)^{2}}\left\{\rho_{0}+f(r)\right\}=\lambda
\end{equation}
Solving (\ref{38}) we have
\begin{equation}\label{40}
\left(\sqrt{\lambda g(t)-1}+\sqrt{\lambda g(t)}\right)^{5}exp\left\{\left(\lambda g(t)-5\right)\sqrt{\frac{\lambda g(t)}{\lambda g(t)-1}}\right\}=exp\left\{\lambda^{\frac{3}{2}}\left(t-t_{0}\right)\right\}
\end{equation}
and
\begin{equation}\label{41}
h(r)=\sqrt{\frac{\rho_{0}+f(r)}{\lambda}}
\end{equation}
where $t_{0}$ appears as the constant of integration. From the solution (\ref{40}), we note that in the asymptotic region(i.e.$t\rightarrow -\infty$), $R(t,r)$ has a finite non-zero value. ($g(t)\rightarrow \frac{1}{\lambda}$). Also, the asymptotic behaviour of $g(t)$ is shown in figure 4.

 Thus, emergent scenario is possible in LTB model with matter only in the form of a cosmological constant.
\begin{figure}
\includegraphics[height=4in, width=4in]{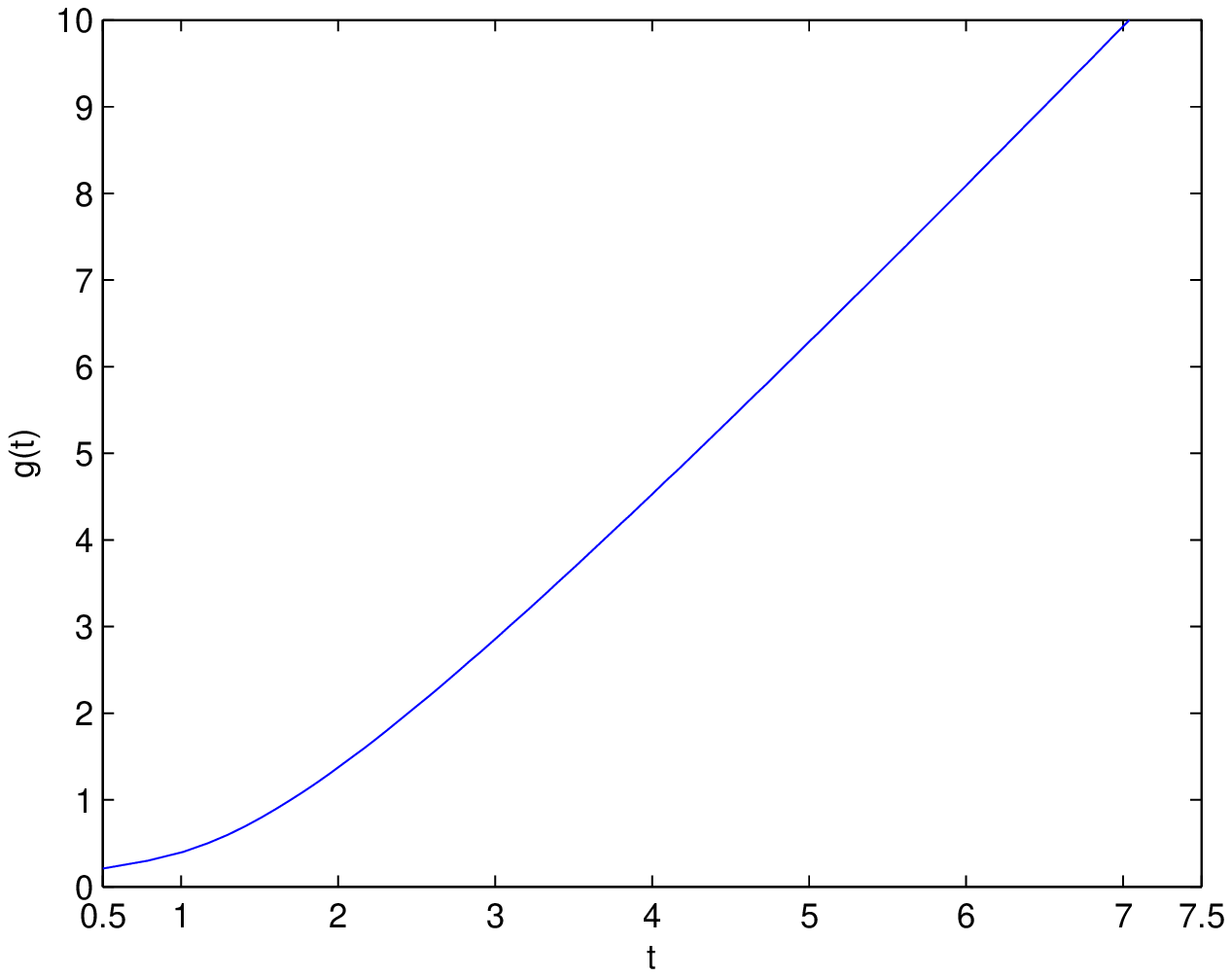}\\
\vspace{1mm} ~~~~~~~~~~~~~~~~~~~~~~~~~~~~~~~~~~~~~~Fig 4\\
\vspace{1mm}
Fig 4 shows the variation of the function  $g(t)$ w.r.t. time t.
\vspace{1mm}
\end{figure}

\section{Discussion : }
In this work, first we have examined the possibility of emergent scenario for general honogeneous and anisotropic model of the universe filled with matter having non-interacting two components -one in the form of perfect fluid with linear equation of state $p=\omega\rho$ and a scalar field (real, phantom or tachyonic) with potential as the other component. For real scalar field and  tachyonic scalar field , when the perfect fluid matter component is not of exotic nature (i.e. does not violate the weak energy condition) then emergent scenario is not possible for flat Bianchi I model $(k=0)$ and open Bianchi III model $(k=-1)$ but for closed Kantowaski-Sachs model $(k=+1)$, emergent scenario is possible when inequality (\ref{17a}) holds. When the perfect fluid matter component is of phantom nature,then emergent scenario is possible for all models in real scalar field and tachyonic scalar field provided restriction (\ref{17b}) holds.
For phantom scalar field and phantom tachyonic scalar field, when the perfect fluid matter component is not of exotic nature, then emergent scenario is always possible for flat and open models while for closed model it is possible when inequality (\ref{17c}) holds. But when the perfect fluid matter component is of phantom nature, then emergent scenario is possible for all three models provided restriction (\ref{17d}) holds. As these restrictions for emergent scenario depend on the anisotropy scalar $\sigma$ so the results are distinct from the isotropic model.

In the second case we have studied the emergent scenario for LTB model. Here it is possible to have a solution assuming the scale factor to be inseparable(product form) and the matter is purely a cosmological constant. The asymptotic analysis shows that in the infinite past there is no singularity and the scale factor grows with time. In future it will be interesting to study the emergent scenario with general form of matter, particularly, with exotic matter.

\begin{contribution}

{\bf Acknowledgement :}

 SM is thankful to Jadavpur University for allowing her to use the
library and laboratory facilities. NM wants to thank CSIR, India for awarding JRF. RB is thankful to West Bengal State Govt for awarding JRF.

\end{contribution}

\end{document}